\documentclass[conference]{IEEEtran}
\IEEEoverridecommandlockouts

\usepackage[utf8]{inputenc}
\usepackage[T1]{fontenc}
\usepackage{amsmath,amssymb,amsfonts}
\usepackage{graphicx}
\usepackage{textcomp}
\usepackage{xcolor}
\usepackage{hyperref}
\usepackage{booktabs}
\usepackage[numbers,sort&compress]{natbib}
\usepackage{tabularx}
\usepackage{booktabs}
\usepackage{array}
\usepackage{ragged2e}
\usepackage{enumitem}
\newcolumntype{Y}{>{\RaggedRight\arraybackslash}X}

\def\BibTeX{{\rm B\kern-.05em{\sc i\kern-.025em b}\kern-.08em
    \sc i\kern-.025em b\kern-.08em \sc i\kern-.025em b\kern-.05em}} 

\begin{document}

\title{Factors Impacting Developer Efficiency:\\Results from an Adaptive Longitudinal Study}

\author{
  \IEEEauthorblockN{Danilo Ribeiro\IEEEauthorrefmark{1}\IEEEauthorrefmark{2},
                    Breno Alves\IEEEauthorrefmark{1},
                    Gabriel
                    Souza
                    \IEEEauthorrefmark{1},
                    César França\IEEEauthorrefmark{1},
                    Alberto Souza\IEEEauthorrefmark{3}}
  \IEEEauthorblockA{\IEEEauthorrefmark{1}Franssa\\
  Recife, Brazil\\
  \{danilo, breno, gabriel, cesar\}@franssa.com}
  \IEEEauthorblockA{\IEEEauthorrefmark{2}CESAR School\\
  Recife, Brazil}
  \IEEEauthorblockA{\IEEEauthorrefmark{3}DevEficiente\\
  São Paulo, São Paulo\\
  alberto@deveficiente.com}
}

\maketitle

\begin{abstract}
\textbf{Context:} Developer efficiency is driven by technical, organizational, and personal factors; however, there is a lack of longitudinal studies exploring how these variables evolve over time.
\textbf{Objective:} This study investigates the primary factors hindering the perceived efficiency of developers within a consulting and professional development context. It analyzes the variance of these factors across recurring data collection cycles, complemented by the developers' qualitative accounts.
\textbf{Method:} We conducted a mixed-methods longitudinal case study applying the Adaptive Developer Efficiency Monitoring Method (ADEMM) to a cohort of 27 external software developers. The data collection comprised twelve waves of periodic surveys and eighteen semi-structured interviews. Quantitative data were evaluated using statistical analysis, while qualitative data underwent thematic analysis.
\textbf{Results:} The most frequent bottlenecks identified were \textit{organizational dependencies} and \textit{waiting for external validation}, which remained structurally stable throughout the monitoring period, followed by technical knowledge gaps, which declined as developers adapted to their work context. A generative AI usage barrier emerged in the qualitative data nine waves into the study and was subsequently incorporated into the survey instrument, becoming the most frequently coded theme in the interviews. The interviews corroborated the quantitative findings, with insufficient requirements documentation and organizational dependencies as the most recurrent themes alongside AI-related challenges.
\textbf{Conclusions:} Perceived developer efficiency is highly dynamic and cannot be accurately captured through a single cross-sectional measurement. Adaptive monitoring via ADEMM successfully identified an emerging factor, generative AI usage barriers, that a fixed instrument defined at the study's onset would have missed entirely, and directly informed a concrete organizational intervention during the study itself. For organizations managing external developers, strategic actions must be directed toward resolving external dependencies, streamlining communication channels, and supporting developers' evolving use of AI tools.
\end{abstract}


\section{Introduction}
\label{sec:introduction}

Software developer efficiency is recognized as a factor that influences software quality, delivery time, and the development experience~\cite{forsgren2018accelerate,space2021,noda2023devex}. However, this efficiency does not depend solely on individual technical ability; it is influenced by organizational, technical, and cognitive factors that vary over time~\cite{meyer2019,chapetta2020towards,razzaq2024systematic}. Barriers such as dependencies between teams, frequent priority changes, or communication difficulties present characteristics distinct from difficulties related to technical knowledge or familiarity with systems, requiring different mitigation strategies~\cite{cheng2022improves,d2024measuring,razzaq2024systematic}. Understanding how these factors evolve during everyday work is, therefore, essential to support more effective organizational interventions.

In recent years, several studies have investigated factors associated with developer productivity, experience, and efficiency, identifying barriers related to infrastructure, collaboration, code quality, interruptions, organizational dependencies, technical knowledge, and the use of development support tools~\cite{meyer2019,canedo2019factors,cheng2022improves,razzaq2024systematic,d2024measuring,coutinho2024role}. Although these studies have significantly broadened the understanding of the factors that influence developers' work, two limitations remain underexplored.

The first concerns the longitudinal tracking of these factors. Most of the literature relies on cross-sectional studies, in which data are collected at a single point in time, making it impossible to distinguish persistent barriers from temporary difficulties~\cite{meyer2019,canedo2019factors}. Longitudinal studies do exist, but they typically employ instruments defined before data collection begins and that are poorly adaptable to changes observed during the investigation~\cite{cheng2022improves,d2024measuring}. As a consequence, factors that emerge over the course of a study, such as those related to the growing use of Generative Artificial Intelligence tools, may not be captured when they were not anticipated at the outset~\cite{coutinho2024role,becker2025measuring}.

The second limitation relates to the organizational context under investigation. Most longitudinal studies follow developers who belong to a single organization and combine survey data with information from repositories, internal tools, or development pipelines~\cite{meyer2019,d2024measuring}. However, there are scenarios in which the organization responsible for the monitoring does not have access to the internal environment of the companies where the developers work, as occurs in consulting firms and professional development programs. In these contexts, understanding the evolution of barriers depends predominantly on the developers' own perceptions, making longitudinal monitoring an additional challenge.

To investigate these gaps, this paper presents the results of a mixed-methods longitudinal study conducted with 27 software developers followed across twelve data collection waves between February and June 2026. The study allowed us to analyze how different factors associated with efficiency evolved during the monitoring period, identifying persistent, adaptive, and emergent patterns observed at different points in the investigation.

The study is guided by the following research questions:

\begin{itemize}

\item \textbf{RQ1.} Which factors most frequently affect developers' perceived efficiency throughout the monitoring period?

\item \textbf{RQ2.} How do these factors vary, persist, or diminish across the recurring data collection waves, and what longitudinal behavioral profiles do they exhibit?

\item \textbf{RQ3.} What themes emerge from developers' qualitative accounts of efficiency barriers, and how do they complement the quantitative findings?

\end{itemize}

This paper contributes to Empirical Software Engineering in three ways. First, it presents evidence that different factors associated with efficiency exhibit distinct longitudinal behavioral profiles, distinguishing persistent, adaptive, and emergent barriers. Second, it shows that assessments conducted at different points in the monitoring period can produce different interpretations of the relevance of the same barriers, highlighting the limitations of purely cross-sectional analyses. Third, it demonstrates how longitudinal monitoring made it possible to identify emerging factors and support an organizational intervention during the course of the study.

The remainder of this paper is organized as follows. Section~\ref{sec:background} reviews the literature on developer efficiency and longitudinal studies. Section~\ref{sec:method} describes the research context, participants, and data collection and analysis procedures. Section~\ref{sec:results} presents the results obtained for the research questions. Section~\ref{sec:discussion} discusses the main findings and their implications. Section~\ref{sec:threats} presents the threats to validity. Finally, Section~\ref{sec:conclusion} concludes the paper and points to directions for future work.

\section{Background}
\label{sec:background}

\subsection{Developer Efficiency}

Productivity and efficiency are related but distinct
concepts~\cite{meyer2014software,coelho2025software}.
Productivity refers to the outcomes achieved, whereas efficiency
emphasizes the relationship between those outcomes and the time, effort, and
resources required to produce them.
In practical terms, efficiency means producing high-quality work with low
friction: completing more tasks does not necessarily indicate higher
efficiency when developers must deal with excessive waiting, rework,
interruptions, or cognitive effort without
payoff~\cite{noda2023devex,coelho2025software}.

The literature recognizes that efficiency is multidimensional and cannot be
captured by a single metric~\cite{forsgren2018accelerate}.
The SPACE framework~\cite{forsgren2018accelerate} organizes productivity
dimensions into satisfaction and well-being, performance, activity,
communication and collaboration, and efficiency and flow.
From the developer experience perspective, feedback loops, cognitive load,
and flow state are the most actionable factors for improving day-to-day
efficiency~\cite{noda2023devex}.
Slow builds, lengthy reviews, insufficient documentation, complex systems,
and organizational dependencies consume time and mental effort without
adding value to the final outcome; fast feedback, clear priorities, and
fewer interruptions allow developers to generate value with less
friction~\cite{noda2023devex,razzaq2024systematic}.

This characterization implies that barriers to efficiency are not static.
They arise, persist, change form, and diminish in response to changes in
the technical and organizational context and in the tools available.
Developers' adaptation to an organization's technical ecosystem, for
example, can progressively reduce knowledge-related barriers over the
first months of a project~\cite{d2024measuring}.
The adoption of Generative Artificial Intelligence tools, in turn,
introduced entirely new categories of friction, such as difficulty
obtaining adequate responses from the models and time spent validating
automatically generated code, which were not observed in the same contexts
before their widespread adoption~\cite{coutinho2024role}.
Monitoring efficiency over time is, therefore, necessary to distinguish
structural and persistent barriers from temporary difficulties associated
with specific phases of the work.

\subsection{Factors Affecting Developer Efficiency}

The literature identifies four main categories of factors affecting
developers' perceived efficiency: technical, organizational, cognitive,
and, more recently, factors related to the use of Artificial Intelligence
tools.

Technical conditions directly influence perceived efficiency.
Code quality, technical debt, infrastructure support, and team
communication have been associated with changes in perceived
productivity~\cite{cheng2022}.
Among these factors, code quality yielded a notable finding: improvements
in perceived code quality tended to precede improvements in perceived
productivity~\cite{cheng2022}.
Understandable and maintainable code reduces the effort required to
understand the system, implement changes, and fix problems, whereas
technical debt and low quality increase task complexity and generate
rework~\cite{cheng2022}.

Organizational conditions affect workflow continuity.
Available resources, task assignment compatible with developers'
experience, and reduced interruptions have been identified as positive
factors for productivity~\cite{razzaq2024systematic}.
In contrast, code complexity, frequent context switching between work
contexts, and lack of standardization increase cognitive
effort~\cite{razzaq2024systematic}.
Organizational conditions also change over time: during the transition to
remote work, for example, access and connectivity emerged as relevant
barriers, and perceived productivity recovered after these conditions
improved~\cite{d2024measuring}.
This result demonstrates that efficiency problems can arise from changes
in context and can be reduced through organizational interventions, which
justifies longitudinal monitoring rather than one-off assessments.

The cognitive impact of interruptions is particularly relevant because it
is not limited to the downtime itself.
Developers need to rebuild the mental context of the task before resuming
work after an interruption, and this additional effort can be more costly
than the interruption itself~\cite{razzaq2024systematic}.
Interruptions arising from external dependencies, waiting for approvals
or validations, and frequent priority changes combine the cost of the
interruption with the cost of uncertainty about what to do next.

The use of Generative Artificial Intelligence tools introduced a fourth
category of factors, still in an early stage of systematic investigation.
A pilot study found that generative AI tools can support development
activities by reducing effort and supporting workflow continuity, but
they also introduce challenges related to the reliability of responses,
prompt crafting, and validation of the generated
outputs~\cite{coutinho2024role}.
These challenges suggest that the benefits of AI depend on the type of
task, the professional role, and developers' ability to evaluate the
responses received, and cannot be generalized to any development
activity.

\subsection{Related Work}

Prior studies have identified and categorized the factors that affect
developer productivity and efficiency.
Canedo and Santos~\cite{canedo2019factors} identified 37 factors
distributed across four groups, namely people, product, organization, and
open-source projects, based on a systematic review combined with a
practitioner survey.
Razzaq et al.~\cite{razzaq2024systematic} reviewed 218 studies and
identified 33 factors related to developer experience and 41 practices
that influence them, showing that adequate resources and fewer
interruptions are consistently associated with positive outcomes.
Meyer et al.~\cite{meyer2019} showed, through a diary study with
developers, that productive days are associated with progress, focus, and
the absence of interruptions, while excessive meetings and fragmented
tasks reduce perceived efficiency.

Longitudinal research is less common, but it provides perspectives that
cross-sectional studies cannot capture.
D'Angelo et al.~\cite{d2024measuring} used recurring surveys to track
developer experience over time, demonstrating that different factors
change at different rates and that a single collection point can capture
an atypical moment.
Kuutila et al.~\cite{kuutila2021individual} combined experience sampling
with repository data over eight months and showed that individual
variance accounts for most of the differences in well-being and
productivity, a result that cross-sectional instruments could not have
revealed.
Russo et al.~\cite{russo2024developers} conducted a two-year study and
showed that changes in working conditions produce systematic variations
in the monitored indicators.

Regarding the use of Generative Artificial Intelligence, Coutinho et
al.~\cite{coutinho2024role} conducted a four-week pilot study and
identified both benefits and challenges related to the use of these tools
in software development.
The small sample and short observation period limit the conclusions that
can be drawn about the persistence and evolution of these effects over
time.

Compared with prior work, this study contributes by longitudinally
tracking developers who work at distinct organizations, with no direct
tie to the organization responsible for the monitoring, across twelve
data collection waves. This context makes it possible to observe whether
the barriers identified in the literature behave persistently or
temporarily when monitored continuously, and whether new dimensions
emerge during the monitoring period, as occurred with the use of
Generative Artificial Intelligence tools in this study. Unlike existing
longitudinal studies, which use fixed instruments defined before data
collection begins~\cite{cheng2022,d2024measuring}, the monitoring
conducted with ADEMM made it possible to incorporate this emergent
dimension without interrupting the continuity of data collection.


\section{Research Method}
\label{sec:method}

This study adopted a mixed-methods longitudinal case study design,
combining recurring surveys and semi-structured interviews to
investigate how factors affecting developer efficiency vary over
time. The monitoring was conducted through the
Adaptive Developer Efficiency Monitoring Method (ADEMM) \cite{ademm_metodo}
, an
adaptive method that combines recurring quantitative data collection,
qualitative interview waves, and iterative instrument revision in
conjunction with the program's organizational stakeholder. The
process of building and refining ADEMM, including the five BIE
cycles (\textit{Build-Intervene-Evaluate}) and the three formalized
design principles, is described in detail in~\cite{ademm_metodo}. This
section describes the context, participants, and data collection and
analysis procedures adopted in this instantiation of the method.
\subsection{Context and Participants}
\label{sec:participants}

The study was conducted between February and June 2026 within the
context of the \textit{Dev Eficiente} course, a professional training
and consulting program offered by an education organization to
software developers affiliated with different client companies.
The contracting organization had no employment relationship with, nor
operational control over, the participants: its interest was to
understand which barriers affected its students' efficiency in order
to guide course content, mentoring, and the program's training
portfolio.

Eligible participants were developers already enrolled in the course,
with no external recruitment or additional referral process.
In total, 27 developers were followed throughout the study, working at
distinct companies and teams, with no organizational tie beyond their
participation in the program.
This heterogeneity makes it possible to observe barrier patterns that
cut across distinct organizational contexts, rather than reflecting the
specific conditions of a single company.
All participants were informed of the research objectives and formally
consented to the collection and use of their data through an Informed
Consent Form.
Table~\ref{tab:seniority} presents the distribution by seniority level.

\begin{table}[ht]
\centering
\caption{Distribution of participants by seniority level.}
\label{tab:seniority}
\begin{tabular}{lcc}
\toprule
\textbf{Seniority level} & \textbf{n} & \textbf{\%} \\
\midrule
Senior &  12 & 44.4 \\
Mid-level  &  10 & 37.0 \\
Junior &   5 & 18.5 \\
\midrule
\textbf{Total} & \textbf{27} & \textbf{100.0} \\
\bottomrule
\end{tabular}
\end{table}

\subsection{Data Collection}
\label{sec:data-collection}

\subsubsection{Recurring surveys}

The study adopted a panel design~\cite{lynn2009methods}, with the same
sample of 27 participants followed across twelve data collection waves.
Surveys were administered weekly between weeks 1 and 7 and biweekly
between weeks 8 and 12, totaling approximately four months of
monitoring.

The instrument combined binary items (Yes/No) covering friction factors
identified in the literature and factors emerging from the data, with
two open-ended questions per wave: one asking participants to describe
the main factor they perceived that week, and another asking them to
indicate anything relevant not covered by the closed items.
Following the ADEMM criteria~\cite{ademm_metodo}, the instrument went
through five versions over the course of the study, labeled Surveys I
through V.
Table~\ref{tab:iterations} presents the correspondence between each
version and the data collection weeks.

\begin{table}[ht]
\centering
\caption{Instrument versions across the twelve data collection waves.}
\label{tab:iterations}
\begin{tabular}{lccc}
\toprule
\textbf{Version} & \textbf{Weeks} & \textbf{Closed items} \\
\midrule
Survey I   & Week 1        & 49 \\
Survey II  & Weeks 2--5    & 19 \\
Survey III & Weeks 6--7    & 17 \\
Survey IV  & Weeks 8--9    & 17 \\
Survey V   & Weeks 10--12  & 16 \\
\bottomrule
\end{tabular}
\end{table}

The sample remained stable in terms of participant identity throughout
the study; there was, however, intermittent non-response across waves.
The number of respondents per week ranged between 13 and 25 out of the
27 participants, a pattern consistent with the literature on
longitudinal panels, which points to occasional non-response as a common
phenomenon in the absence of definitive attrition~\cite{lynn2009methods}.

Percentage frequencies of the reported factors were calculated using the
fixed denominator of the 27 enrolled participants, rather than the
variable number of respondents in each wave.
This convention operationalizes each percentage as a proportion of the
full cohort, making values directly comparable across weeks with
different response rates.
As a consequence, weeks with higher non-response produce conservatively
lower percentages for the corresponding items.

\subsubsection{Semi-structured interviews}

The qualitative channel operated across two waves of interviews
conducted over the course of the monitoring period.
The first wave, comprising eight interviews, investigated general
factors related to work efficiency, including task organization,
requirements, estimation, and the tools used.
Analysis of this wave identified two themes that were not adequately
covered by the closed items then in use: the use of Generative
Artificial Intelligence tools and process-related barriers to workflow.
The second wave, comprising ten interviews, delved specifically into
these two themes.
In total, 18 of the 27 participants were interviewed.
Interviews were recorded, transcribed, and analyzed by the researchers.

For the theme of process-related barriers to workflow, the decision, made
jointly with the program stakeholder, differed from the one adopted for
the use of Artificial Intelligence. It was determined that the closed
items already present in the instrument, such as priority changes, lack
of clarity in technical priorities, and rework due to misalignment, were
already sufficient to quantitatively capture the incidence of the
problem across waves. Deepening this theme therefore did not require
creating new closed items, and was instead pursued entirely through the
qualitative channel, using the revised interview script applied in the
second wave of interviews.

\subsection{Data Analysis}
\label{sec:analysis}

\subsubsection{Quantitative data}

Survey data were analyzed using descriptive statistics, including
percentage frequencies of the reported factors in each data collection
wave.
This analysis made it possible to observe variations, persistence, and
reductions in the occurrence of factors over time, as presented in
Section~\ref{sec:results}.
Direct longitudinal comparisons were restricted to items that remained
textually stable across consecutive instrument versions.
For items that were removed, reworded, or introduced later, the results
were interpreted as evidence situated within the period in which they
were present.

\subsubsection{Qualitative data}

Interview data were analyzed using deductive and inductive thematic
analysis~\cite{braun2006thematic}.
The deductive analysis was guided by the factors investigated in the
surveys and by the interview script topics; the inductive analysis made
it possible to identify emergent patterns in the accounts that were not
anticipated in the quantitative instruments.
The process followed the steps of complete and repeated reading of the
transcripts, coding of relevant excerpts, iterative grouping of codes
into provisional categories, and revision of categories as new
interviews were analyzed.
In the end, the 320 coded excerpts from the 18 interviews were organized
into 236 distinct codes, subsequently grouped into seven analytical
themes, presented in Section~\ref{sec:results}.

\subsubsection{Triangulation}

Triangulation between the quantitative and qualitative data occurred at
two points.
During the monitoring period, patterns identified in the analysis of
each interview wave were compared with the percentages observed in the
corresponding surveys and presented to the program stakeholder in
review meetings, guiding revisions to the instrument.
At the end of the study, the seven analytical themes from the interviews
were systematically compared against the most frequent factors from the
surveys, as described in Section~\ref{sec:results}.
This triangulation made it possible to verify whether the barriers
identified by each collection channel converged on the same patterns.

\section{Results}
\label{sec:results}

\begin{figure}[t]
  \centering
  \includegraphics[width=\linewidth]{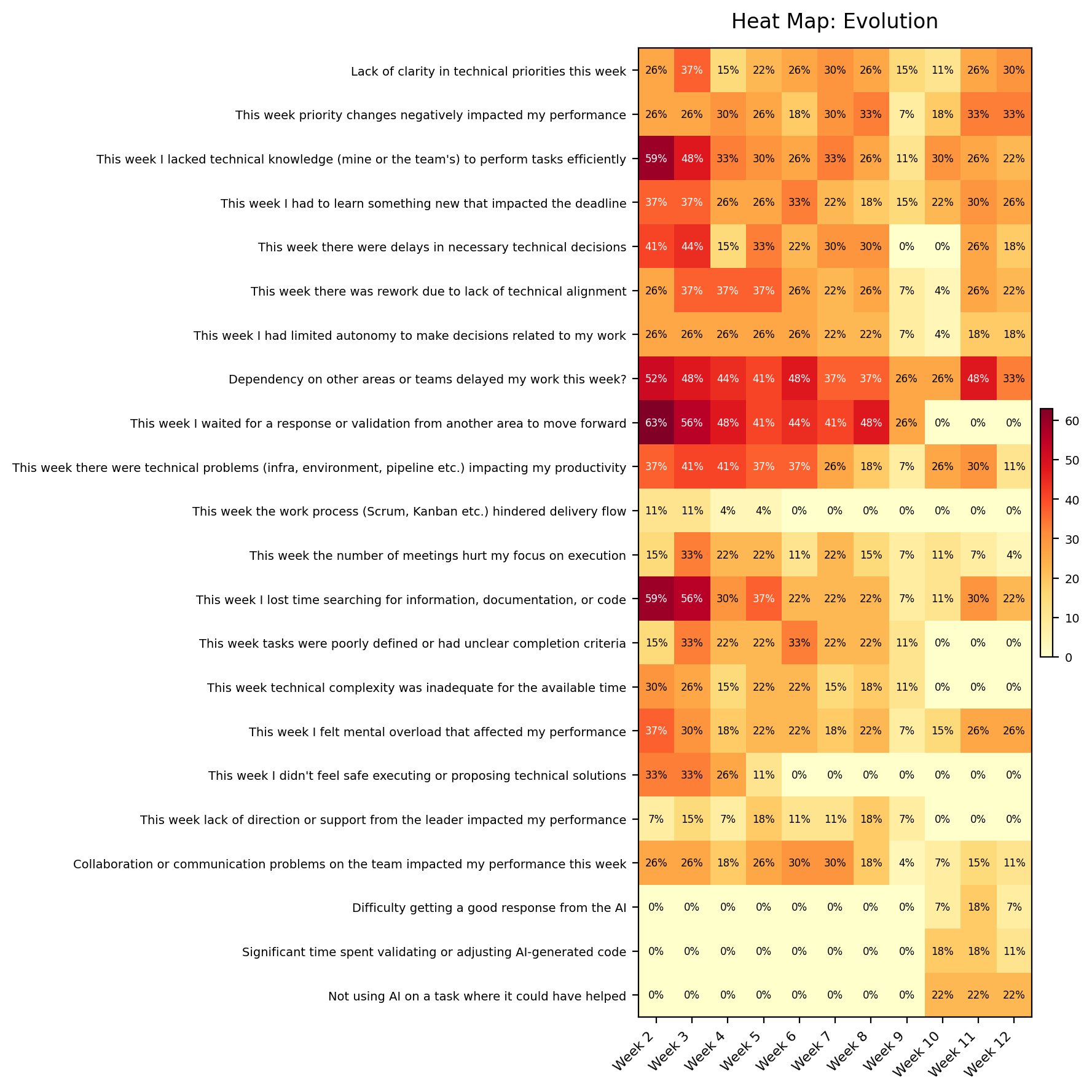}
  \caption{Percentage frequency of the reported factors by data
  collection wave (Weeks~2 to~12). Cells with 0\% indicate that the
  item was not present in the instrument that week.}
  \label{fig:heatmap}
\end{figure}

This section presents the results obtained through the recurring
surveys and the semi-structured interviews, organized by the three
research questions. As an overview of the longitudinal monitoring,
Figure~\ref{fig:heatmap} summarizes the percentage frequency of each
factor across all data collection waves, making it possible to
visualize the persistence, variation, and emergence of factors
throughout the study. Detailed analyses of these behaviors are
presented in the answers to the research questions. Cells with a value
of 0\% indicate that the item was not present in the instrument that
week.

\subsection{RQ1 — Which factors most frequently affect developers'
perceived efficiency}
\label{sec:rq1}

The most frequently reported factors throughout the monitoring period
centered on organizational dependencies and coordination difficulties.
The item \textit{Dependency on other areas or teams delayed my work
this week} showed percentages between 25.9\% and 51.9\% across all
waves analyzed.
Similarly, the item \textit{This week I waited for a response or
validation from another area in order to move forward} recorded
frequencies between 25.9\% and 63.0\% in the waves in which it was
present.
These two items concentrated the highest percentages observed
throughout the monitoring period.

Next, factors related to knowledge gaps and information-seeking were
frequent.
The item \textit{This week I lacked the technical knowledge to carry
out tasks efficiently} ranged between 11.1\% and 59.3\%, with higher
values in the initial weeks.
The item \textit{This week I lost time searching for information,
documentation, or code} showed percentages between 7.4\% and 59.3\%.
Technical infrastructure, environment, or pipeline problems were
reported between 7.4\% and 40.7\%.

Factors related to work organization also appeared recurrently.
Priority changes ranged between 7.4\% and 33.3\%, lack of clarity in
technical priorities between 11.1\% and 37.0\%, and rework due to
misalignment between 3.7\% and 37.0\%.
Delays in technical decisions peaked at 44.4\% in Week~3 and dropped to
0\% in Weeks~9 and~10, before partially recovering to 25.9\% in
Week~11.
Factors related to individual experience, such as mental overload
(7.4\%–37.0\%) and an excessive number of meetings (3.7\%–33.3\%), were
also reported with regularity.

Factors related to the use of Generative Artificial Intelligence were
assessed only in the three final waves, totaling $n = 55$ valid
responses.
The item \textit{Not using AI on a task where it could have helped}
showed a stable frequency of 22.2\% in Weeks~10, 11, and~12.
The item \textit{Significant time spent validating or adjusting
AI-generated code} ranged between 11.1\% and 18.5\%, and
\textit{Difficulty getting a good response from the AI} between 7.4\%
and 18.5\%.

\subsection{RQ2 - How do those factors vary, persist, or resolve across recurring survey cycles?}
\label{sec:rq2}

\begin{figure*}[t]
  \centering
  \includegraphics[width=\textwidth]{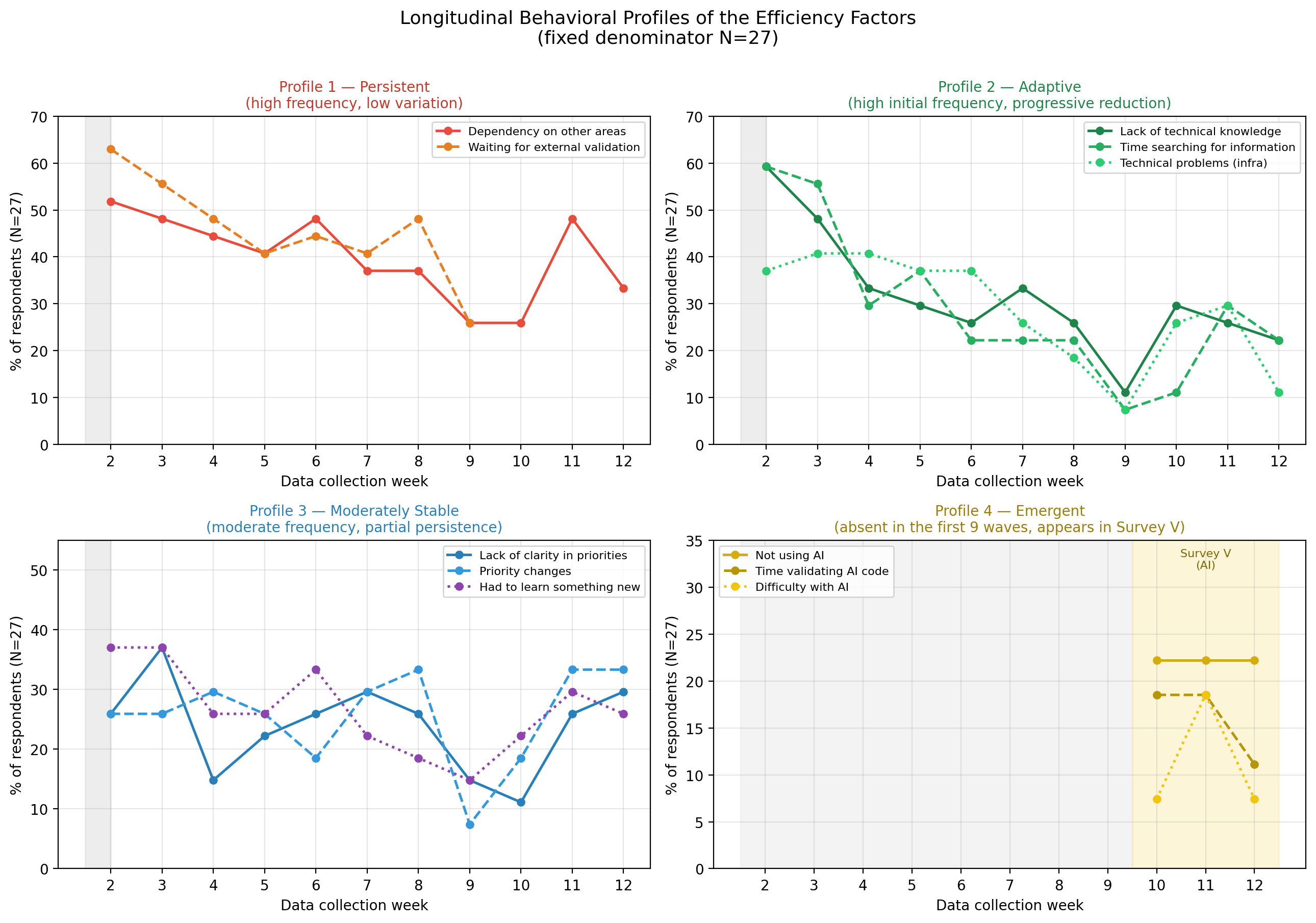}
  \caption{Longitudinal behavioral profiles of the efficiency factors
  across the twelve data collection waves (fixed denominator $N=27$).
  Profile~1 (persistent): high mean frequency and low weekly variation,
  indicating a structural barrier.
  Profile~2 (adaptive): high initial frequency with a significant
  progressive decline ($r \leq -0{,}75$; $p < 0{,}01$), compatible with
  adaptation to the technical context.
  Profile~3 (moderately stable): moderate frequency with no clear
  trend, present throughout the monitoring period.
  Profile~4 (emergent): absent in the first nine waves and incorporated
  into the instrument in Survey~V after being identified in the
  qualitative data.}
  \label{fig:perfis}
\end{figure*}

The longitudinal analysis of the factors revealed four distinct
behavioral profiles, determined by the combination of mean frequency,
coefficient of variation, and trend across waves. In this paper, the
term \textit{profile} refers to the longitudinal behavior exhibited by
a factor across the data collection waves, not to a profile of
participants or groups of developers.
Figure~\ref{fig:perfis} presents the profiles graphically, and
Table~\ref{tab:perfis} characterizes them with the corresponding
metrics.

\begin{table}[t]
\centering
\caption{Longitudinal behavioral profiles of the efficiency factors.
CV = coefficient of variation; slope = trend line slope in percentage
points per week.}
\label{tab:perfis}
\footnotesize
\renewcommand{\arraystretch}{1.2}
\begin{tabularx}{\linewidth}{
>{\RaggedRight\arraybackslash}p{1.5cm}
>{\RaggedRight\arraybackslash}p{3.0cm}
Y}
\toprule
\textbf{Profile} &
\textbf{Representative factors} &
\textbf{Characterization} \\
\midrule

Persistent &
Dependency on other areas (Q8);
waiting for external validation (Q9) &
Mean 40--46\%; CV 0.23--0.24 (low);
slope not significant with the variable denominator
($r = -0{,}14$; $p = 0{,}68$).
Structural barrier, does not respond to accumulated technical
experience. \\

\midrule

Adaptive &
Lack of technical knowledge (Q3);
time spent searching for information (Q13);
technical infrastructure problems (Q10) &
Mean 28--31\%; CV 0.41--0.56 (medium);
significant negative slope ($r \leq -0{,}75$; $p < 0{,}01$).
Reduction compatible with adaptation to context; reappears when
new demands or technologies emerge. \\

\midrule

Moderately stable &
Lack of clarity in priorities (Q1);
priority changes (Q2);
rework due to misalignment (Q6);
mental overload (Q16) &
Mean 20--27\%; CV 0.27--0.45 (low-medium);
no statistically significant trend.
Consistently present throughout the monitoring period at a moderate
intensity. \\

\midrule

Emergent &
Not using AI (Q22);
time spent validating AI-generated code (Q21);
difficulty with AI (Q20) &
Absent in Weeks~1--9; incorporated into the instrument in Survey~V
after being identified in the qualitative data.
CV = 0.00 for Q22 (22.2\% stable across the three final waves). \\

\bottomrule
\end{tabularx}
\end{table}


The \textbf{persistent} profile brings together the two factors with the
highest mean frequency in the study and the lowest weekly variability:
dependency on other areas (mean 40.1\%; CV = 0.23) and waiting for
external validation (mean 45.8\%; CV = 0.24).
The linear correlation analysis using the fixed denominator of 27
participants indicates a formally significant downward trend
($r = -0{,}63$; $p = 0{,}038$), but this reduction coincides with
Weeks~9 and~10, the period with the lowest panel response rates (13 and
17 respondents, respectively).
When the correlation is recalculated using the effective number of
respondents as the denominator, the trend is no longer significant
($r = -0{,}14$; $p = 0{,}68$), and when participation recovered in
Week~11, dependency returned to 48.1\%, close to the initial value of
51.9\% in Week~2.
These factors reflect organizational conditions at the participants'
employer companies and do not respond to the accumulation of individual
technical experience.

The \textbf{adaptive} profile shows the opposite pattern: high frequency
in the early waves, a statistically significant progressive decline
over the course of the study, and a partial rebound in the final weeks.
Lack of technical knowledge dropped from 59.3\% in Week~2 to 11.1\% in
Week~9 ($r = -0{,}75$; $p = 0{,}007$), but rose again to 29.6\% in
Week~10 and 22.2\% in Week~12.
An identical pattern was observed for time spent searching for
information, documentation, or code (59.3\% in Week~2; 7.4\% in Week~9;
29.6\% in Week~11) and for technical infrastructure problems (40.7\% in
Weeks~3--4; 7.4\% in Week~9; 25.9\% in Week~10).
This behavior is compatible with a process of adaptation to the
technical ecosystem, codebase, and business rules of the employer
organizations, but the reduction is not permanent: new demands, project
changes, or increased complexity reintroduce barriers that appeared to
have been overcome.

The \textbf{moderately stable} profile groups factors with a mean
frequency between 20\% and 27\% over the course of the monitoring
period, although the individual weekly values fluctuate across a wider
range, with no statistically significant trend but with consistent
presence.
These are barriers that exist continuously but at an intensity that
does not dominate the overall picture: lack of clarity in priorities,
priority changes, rework due to misalignment, having to learn something
new, limited autonomy, and mental overload.
This profile indicates recurring working conditions that do not come to
dominate participants' perceptions, but that also do not disappear.

The \textbf{emergent} profile is the most relevant to this study's central
argument.
The three items related to the use of Generative Artificial Intelligence
did not exist in the instrument for nine waves because they were not
anticipated at the outset of the study.
They were first identified in the qualitative data from the first
interview wave, conducted between Weeks~2 and~5, and were incorporated
into Survey~V in Weeks~10 through~12.
The item \textit{Not using AI on a task where it could have helped}
showed CV = 0.00, meaning it occurred in exactly 22.2\% of responses
across the three waves in which it was measured.
A fixed instrument defined at the start of the study would have had no
item on generative AI in the final waves, when this theme was the most
recurrent one in the interviews (31 coded excerpts) and had already
prompted a concrete organizational intervention.

Taken together, the four profiles show that the factors affecting
developers' perceived efficiency do not constitute a homogeneous set
that can be monitored with a single static instrument.
Organizational factors are structurally persistent and require
intervention in the workflows and processes of the employer
organizations.
Technical factors are adaptive and respond to documentation, onboarding,
and knowledge-sharing practices, but return when the context changes.
Emergent factors, by definition, can only be captured by an instrument
with a built-in revision mechanism during data collection: no fixed
instrument defined before a period of intense technological adoption
can anticipate the dimensions of friction that this adoption will
introduce.

\subsection{RQ3 - What themes emerge from developers' open-ended accounts of efficiency
barriers, and how do they converge with the quantitative findings?}
\label{sec:rq3}

The 18 semi-structured interviews produced 320 coded excerpts,
resulting in 236 distinct codes grouped into seven analytical themes.
Table~\ref{tab:themes} presents these themes and their descriptions.

\begin{table}[t]
\centering
\caption{Analytical themes identified in the semi-structured interviews.}
\label{tab:themes}
\footnotesize
\renewcommand{\arraystretch}{1.15}
\begin{tabularx}{\linewidth}{>{\RaggedRight\arraybackslash}p{3.0cm} Y}
\toprule
\textbf{Theme} & \textbf{Description} \\
\midrule
AI use as support and source of challenges &
Experiences using AI tools in understanding, implementation,
debugging, and decision-support activities. \\
Understanding the business, system, and requirements &
Difficulties related to business rules, existing systems,
requirements, and the information needed to develop a solution. \\
Dependencies, communication, and collaboration &
The need for interaction, alignment, or support from other people,
teams, or areas in order to carry out activities. \\
Organizational processes, planning, and prioritization &
Organization of activities, definition of priorities, effort
estimation, and practices that guide the work. \\
Technical knowledge &
Mastery of technologies, tools, components, and architectural
decisions needed to carry out tasks. \\
Interruptions and focus at work &
Situations that interrupt reasoning, hinder concentration, or
increase mental load. \\
Quality, validation, and rework &
Practices related to testing, code review, delivery validation, and
fixing problems. \\
\bottomrule
\end{tabularx}
\end{table}

The most recurrent theme was AI use as a support and a source of
challenges, with 31 coded excerpts.
Participants reported that AI tools hindered efficiency when used with
insufficient context, when they produced inconsistent suggestions, or
when they required multiple iterations to understand the problem.
One participant stated: ``What I've noticed is that if I ask a few
questions or give a more generic scope... usually the AI's help gets in
the way more than it helps'' (P1).\footnote{Interview quotes were
originally collected in Portuguese and have been translated into
English for this manuscript.}
Another reported that, in diagnostic activities, the responses were
often incomplete: ``Sometimes it gives an answer that works, sometimes
it's incomplete... I keep asking follow-up questions to give it more
context'' (P2).
These accounts reveal that the effort of validating and refining AI
responses is itself a barrier to efficiency, even when the tool is used
with the intention of reducing effort.

The second most frequent theme by volume of excerpts was understanding
the business, system, and requirements, with 29 excerpts.
Participants highlighted insufficient documentation, unclear acceptance
criteria, and legacy systems that were difficult to understand.
One participant summarized: ``The information arrives very vague and
creates unnecessary work, since developers have to figure everything
out for themselves'' (P3).
Another reported that the available documentation frequently became a
bottleneck because it was incomplete or difficult to interpret (P4).

The third theme was dependencies, communication, and collaboration,
with 28 excerpts.
One participant described a case in which they depended on another
professional to obtain access that, had they had it themselves, would
have let them finish the same day: ``If I had had access, I would have
finished it the same day, but since I depended on him and his
availability, it took me a week to finish everything'' (P5).
Communication problems within the team itself were also reported: ``We
have a lot of problems with unclear communication on the team.
Sometimes the team gets in the way more than it helps, because everyone
understands things differently'' (P6).

The four remaining themes accounted for smaller volumes of excerpts:
organizational processes (25), technical knowledge (13), interruptions
and focus (11), and quality, validation, and rework (11).
Accounts of planning described frequent priority changes and urgent
demands that interrupted ongoing work.
Accounts of technical knowledge showed that gaps in specific
technologies or components increased the time needed to complete tasks.
Accounts of interruptions showed that resuming one's train of thought
after consecutive meetings takes significant time, even when the
meetings themselves are necessary.

\subsubsection{Convergence between qualitative and quantitative data}

Table~\ref{tab:convergence} summarizes the correspondence between the
interview themes and the survey factors, with the absolute record
counts from the quantitative data.

\begin{table}[htbp]
\centering
\caption{Convergence between qualitative themes and survey data}
\label{tab:convergence}
\footnotesize
\renewcommand{\arraystretch}{1.3}
\begin{tabularx}{\linewidth}{>{\RaggedRight\arraybackslash}p{3.8cm} >{\RaggedRight\arraybackslash}X}
\toprule
\textbf{Interview theme} & \textbf{Factors and records (N=27)} \\
\midrule
\textbf{Dependencies, communication, and collaboration} \newline (28 excerpts) & 
Dependency on other areas: 119 rec. \newline 
Waiting for a response or validation: 99 rec. \newline 
Collaboration problems: 57 rec. \\
\midrule
\textbf{Organizational processes, planning, and prioritization} \newline (25 excerpts) & 
Priority changes: 76 rec. \newline 
Lack of clarity in priorities: 71 rec. \newline 
Workflow process: 8 rec. \\
\midrule
\textbf{Understanding the business, system, and requirements} \newline (29 excerpts) & 
Searching for information/documentation: 86 rec. \newline 
Poorly defined tasks: 49 rec. \\
\midrule
\textbf{Technical knowledge} \newline (13 excerpts) & 
Lack of technical knowledge: 93 rec. \newline 
Technical/infrastructure problems: 84 rec. \\
\midrule
\textbf{AI use as support and challenges} \newline (31 excerpts) & 
Difficulty with AI responses: 9 rec. \newline 
Time validating generated code: 13 rec. \\
\bottomrule
\end{tabularx}
\end{table}

The strongest convergence occurred for the theme of dependencies,
communication, and collaboration.
In the interviews, participants described delays resulting from waiting
for access, approvals, responses, and validations from other teams.
Consistently, the surveys recorded 119 occurrences of dependency on
other areas and 99 records of waiting for a response or validation,
confirming that external dependency is the most frequent and persistent
barrier in the set.

The themes of understanding the business and technical knowledge also
showed strong convergence.
The qualitative accounts revealed insufficient documentation, unclear
requirements, legacy systems, and technical gaps; the surveys recorded
86 cases of time spent searching for information and 93 occurrences of
lack of technical knowledge.

The AI use theme showed the weakest numerical convergence, with only 9
and 13 quantitative records against 31 qualitative excerpts.
This suggests that the binary survey items captured only a limited part
of the difficulties with AI: the interviews revealed broader barriers,
such as the need to provide adequate context, inconsistency in
responses, and validation effort, which do not translate directly into
a binary presence-or-absence item.
The partial convergence in this theme illustrates the complementary role
of the two data collection channels: the qualitative channel identified
the dimension before a corresponding closed item existed, and the items
included in Survey~V captured only the most observable manifestations
of the phenomenon.

Overall, the qualitative results reinforce the quantitative findings by
showing that the most frequent barriers in the surveys also emerged
spontaneously in the interviews.
This convergence strengthens the evidence that organizational
dependencies, understanding of requirements, technical gaps, and the
challenges of AI use are relevant factors for understanding developers'
perceived efficiency in software development.

\section{Discussion}
\label{sec:discussion}

The four longitudinal behavioral profiles identified in
Section~\ref{sec:rq2} reveal that the barriers to developers' perceived
efficiency do not form a homogeneous set: they have distinct causal
mechanisms, respond to distinct interventions, and require distinct
monitoring strategies.
This typology is the study's main interpretive contribution.
The following subsections discuss each profile in light of the
literature, identify what this study adds to what was already known,
and derive specific implications for research and practice.

\subsection{The persistent profile: barriers that experience does not resolve}

The persistent profile, represented by dependency on other areas and
waiting for external validation, is characterized by high mean
frequency and low weekly variability throughout the entire monitoring
period.
The coefficient of variation analysis (CV = 0.23--0.24) confirms that
these factors do not fluctuate with weekly working conditions: they are
structurally present regardless of who is responding or which phase of
the project is underway.

This result aligns with the literature that identifies organizational
dependencies and validation workflows as determinants of efficiency
that transcend individual effort~\cite{razzaq2024systematic,
d2024measuring}.
What this study adds is longitudinal evidence that these barriers do
not respond to developers' accumulated technical experience: while the
adaptive factors declined progressively in the early weeks,
organizational dependencies remained stable, immune to the technical
learning curve observed in the other profiles.
This dissociation, which is only visible in longitudinal data collected
over a sufficient period, has a direct implication: technical training
initiatives are insufficient when the main blockers depend on approval
cycles, the availability of other teams, and communication workflows
that are outside the developer's control.

The context of this study reinforces this interpretation.
Because the contracting organization has no operational control over
the participants' employer companies, the interventions available to
the \textit{Dev Eficiente} program do not reach the structural causes of
this profile.
What the monitoring revealed is precisely the limit of what
professional education can resolve: organizational coordination
barriers need to be addressed by the employer organizations themselves,
through fewer approval steps, increased autonomy, and more agile
response channels between teams.

\subsection{The adaptive profile: barriers that experience reduces but does
not eliminate}

The adaptive profile, composed of factors related to technical
knowledge, information seeking, and infrastructure, shows a
significant negative slope ($r \leq -0{,}75$; $p < 0{,}01$), indicating
progressive reduction across waves.
This pattern is compatible with a process of adaptation to the
technical ecosystem, codebase, and business rules of the employer
organizations: as participants accumulate familiarity with systems,
tools, and processes, the effort required to locate information and
understand tasks decreases.

The literature suggests that documentation practices, structured
onboarding, and knowledge sharing among developers support this
adaptation process~\cite{razzaq2024systematic}.
What this study's data add is the nonlinear behavior of the reduction:
factors in the adaptive profile do not disappear, they only decrease
temporarily.
The partial rebound in the final weeks, with lack of technical
knowledge returning to 29.6\% in Week~10 after reaching 11.1\% in
Week~9, indicates that new demands, project changes, or increased
complexity reactivate barriers that appeared to have been overcome.
Knowledge support practices therefore need to keep pace with the
evolution of the technical context over time, not just the initial
onboarding period.

This pattern concretely illustrates the risk of cross-sectional
assessments pointed out in the literature~\cite{d2024measuring}.
A data collection conducted exclusively in Week~2 would produce the
interpretation that lack of technical knowledge is the group's dominant
barrier.
A data collection conducted in Week~9 would produce the interpretation
that this factor had been largely resolved.
Both interpretations would be incorrect: the first for failing to
distinguish an adaptation phase from a structural barrier; the second
for confusing a temporary reduction with a definitive elimination.
Only longitudinal monitoring makes this distinction possible.

\subsection{The emergent profile: the study's strongest methodological
argument}

The emergent profile is the most relevant to this paper's central
argument and the one that most clearly justifies the use of an adaptive
instrument instead of a fixed one.

The three items related to the use of Generative Artificial Intelligence
did not exist in the instrument for nine waves because they were
unknown at the start of the study.
They were not anticipated in the literature review that informed the
first version of the instrument, were not mentioned by participants in
the early surveys, and were not part of any of the twelve factor groups
in Survey~I.
They arose spontaneously in the interviews from the first qualitative
wave, conducted between Weeks~2 and~5, as a source of difficulties that
the closed items then in use did not capture.
They were incorporated into the instrument only in Survey~V, in
Weeks~10 through~12.

A fixed instrument defined in February 2026 would have had zero items
on generative AI in the final waves, precisely when this theme was the
most recurrent one in the interviews (31 coded excerpts, more than any
other theme) and when the monitoring had already prompted the
contracting organization to create a new course.
This claim is not hypothetical: it is what every longitudinal study
with a fixed instrument identified in the
literature would have
done~\cite{russo2024developers,kuutila2021individual,d2024measuring}.
The emergent profile, by definition, does not exist before it emerges.

The data on the three AI items also raise a relevant interpretive
question.
Not using the tool on tasks where it could have helped (22.2\%, stable
across the three waves) exceeded the technical difficulty in obtaining
good responses (7.4\%--18.5\%).
This result admits two complementary interpretations: the first is that
some participants had not yet developed the ability to recognize
situations in which AI adds value; the second, supported by recent
evidence, is that non-use may reflect an implicit judgment that the
cost of using the tool outweighs its benefits for certain tasks, given
that AI tools can increase the time required to complete activities
that demand significant prompting, validation, and correction
effort~\cite{becker2025measuring}.
Distinguishing between these two situations requires data on task type
and the reasons for non-use, which the current binary instrument does
not capture.
This finding suggests that AI monitoring items need more granular
formats than yes/no.

\subsection{Qual-quant convergence and the complementary role of the channels}

The correspondence between the interview themes and the survey factors,
documented in Table~\ref{tab:convergence}, confirms that the two
channels identified the same barriers through independent paths.
Across the three profiles of already-known factors, the convergence is
direct and numerically expressive: dependencies and communication sum
to 275 records in the surveys, understanding the business and technical
knowledge together sum to 312 (135 and 177, respectively), and planning
factors sum to 155.

The case of greatest interest is the emergent profile, where the
convergence is numerically smaller (22 survey records against 31
interview excerpts) but structurally more informative.
This divergence does not reflect the absence of the phenomenon: it
reflects the fact that binary items capture only the most observable
manifestations of a barrier, while the interviews capture the full
range of situations that barrier can take.
AI introduces difficulties around insufficient context, inconsistency
of responses, validation effort, and uncertainty about when to use the
tool, none of which translates directly into a presence-or-absence
item.
This divergence reinforces the need to keep the qualitative channel
active throughout monitoring: it was this channel that identified the
emergent profile before any corresponding closed item existed.

\subsection{Implications for research}

The four profiles suggest that studies on developer efficiency should
structure their longitudinal analyses around the distinction between
barrier types, not merely a listing of factors.
The relevant question is not ``which factors appeared,'' but ``what
kind of mechanism sustains each factor over time.''
Persistent factors require variability analysis and denominator
sensitivity testing; adaptive factors require trend analysis with a
rebound test; emergent factors require an active qualitative channel
with an instrument revision mechanism.

This distinction has consequences for the design of future studies.
Instruments and observation windows calibrated to capture short-term
adaptive barriers may be insufficient to document persistent barriers
that manifest at a stable frequency over months.
And fixed instruments, regardless of how long the observation window
is, are inherently unable to capture emergent profiles by definition.
In periods of intense technological change, such as the current spread
of generative AI tools in software development, the cost of fixed
instruments is especially high.

\subsection{Implications for practice}

For factors in the \textbf{persistent profile}, the intervention needs to
reach the developers' employer organizations: reducing approval steps,
increasing autonomy to proceed without waiting for external validation,
and structuring more agile communication channels between teams.
Training and consulting programs that lack access to these
organizations can guide participants in developing workaround
strategies, but they do not eliminate the structural cause of the
barrier.

For factors in the \textbf{adaptive profile}, the most effective
intervention is one that reduces the time needed to adapt to the
technical context: accessible documentation, structured onboarding, and
active knowledge-sharing practices within teams.
The results indicate that these practices need to keep pace with
project changes over time, not just the initial integration period.

For the emergent profile, the most important practical implication was
demonstrated during the study itself: when the monitoring identified
AI use as an emergent barrier, the contracting organization created a
new course focused on requirements analysis and improvement,
responding to the blocking pattern the data revealed.
Similarly, the qualitative deepening of the theme of process-related
barriers to workflow, pursued without creating new closed items through
a decision made jointly with the program stakeholder, also resulted in
a concrete action: the creation of a course to guide students on how to
avoid workflow problems in their activities.
For AI use specifically, the preliminary results suggest that training
initiatives should prioritize developing the ability to recognize when
AI adds value, rather than focusing exclusively on technical prompt
training.

Table~\ref{tab:summary} summarizes the central findings and their
implications, organized by profile.

\begin{table}[t]
\centering
\caption{Summary of findings by longitudinal behavioral profile.}
\label{tab:summary}
\footnotesize
\renewcommand{\arraystretch}{1.2}
\begin{tabularx}{\linewidth}{
>{\RaggedRight\arraybackslash}p{1.6cm}
>{\RaggedRight\arraybackslash}p{2.3cm}
Y}
\toprule
\textbf{Profile} &
\textbf{Key finding} &
\textbf{Implication} \\
\midrule

Persistent &
Frequency stable above 40\% for 8--11 waves; does not respond to
accumulated experience &
Requires changes in the organizational processes of the employer
companies; technical training does not reach the structural cause. \\

\midrule

Adaptive &
Reduction from 59\% to 11\% in the early weeks; rebound to 22--30\%
in the final weeks &
Documentation and knowledge sharing reduce adaptation time, but need
to keep pace with context changes, not just the initial onboarding. \\

\midrule

Moderately stable &
Present between 20--27\% throughout the monitoring period with no
clear trend &
Recurring low-intensity barriers; continuous monitoring is needed to
detect any eventual escalation. \\

\midrule

Emergent &
Absent in the first 9 waves; identified via interviews and
incorporated in Survey~V; stable at 22.2\% in the final 3 waves &
A fixed instrument would have had zero data on generative AI in the
waves where it was the most cited barrier in the interviews. \\

\bottomrule
\end{tabularx}
\end{table}
\section{Threats to Validity}
\label{sec:threats}

Following the framework of Wohlin et al.~\cite{wohlin2012experimentation},
this section discusses threats to construct, internal, external, and
reliability validity.

\subsection{Construct Validity}

The instrument measures self-reported perceptions of efficiency, not
objective productivity measures.
The results should be interpreted as factors perceived by participants
as influential to their efficiency, rather than as direct indicators of
performance or of software development process outcomes.
The interviews complemented the surveys by detailing the context of the
reported factors, but they also depend on the interpretation of both
participants and researchers.

The fixed-denominator convention of 27 participants, rather than the
effective number of respondents per wave, operationalizes each
percentage as a proportion of the full cohort.
This decision increases comparability across weeks but produces
conservatively lower percentages in weeks with higher non-response.
The impact of this convention was examined directly in the analysis of
dependency on other areas, where the correlation with week number
ceased to be significant when the fixed denominator was replaced with
the effective number of respondents.

The instrument was refined across five iterations.
Items added in later versions, such as those related to the use of
Artificial Intelligence, cannot be directly compared across the full
twelve-week period and were interpreted only within the waves in which
they were present.
Direct longitudinal comparisons were restricted to items that remained
textually stable across consecutive iterations.

\subsection{Internal Validity}

The main threat to internal validity stems from variation in
participation across the data collection weeks.
The number of respondents ranged between 13 and 25 out of 27
participants, which may introduce non-response bias: participants who
were more overloaded or more affected by particular barriers may have
been less likely to respond in specific weeks.
This risk is especially relevant in weeks with lower adherence, such as
Week~9, when some of the sharpest reductions in the monitored factors
also occurred.

The study did not include a control group, a counterfactual condition,
or a static instrument for comparison.
The observed variations may be associated with changes in work demands,
in participants' context, in the composition of respondents, or in the
instrument itself, without it being possible to isolate any factor
causally.
It is not possible to causally attribute the reduction in technical
barriers to accumulated experience, to changes in tasks, or to both
simultaneously.

The temporal traceability of the interviews relative to the data
collection waves was not preserved with sufficient precision to allow
direct cross-referencing between qualitative findings from a specific
interview and the percentages for the corresponding week.
This limitation restricts triangulation to aggregate patterns, rather
than specific events.

\subsection{External Validity}

The study was conducted with 27 participants from a single professional
training program in Brazil, affiliated with a single education and
consulting organization.
The results regarding the factors affecting perceived efficiency cannot
be statistically generalized to other groups of developers, companies,
countries, or software domains.

The context investigated has a characteristic that distinguishes it
from most existing longitudinal studies: the organization responsible
for the monitoring had no employment relationship with, nor operational
control over, the participants.
The findings may not replicate at the same magnitude in studies
conducted within a single company, where the researcher has direct
access to the infrastructure, processes, and organizational leadership.

Items related to the use of Artificial Intelligence were monitored only
in the three final weeks.
The results regarding AI are preliminary and do not support
generalizations about long-term trends, persistence, or the temporal
evolution of these factors.

\subsection{Reliability}

The iterative adaptation of the instrument may have reduced
comparability across some waves.
To mitigate this risk, trend analyses were restricted to items that
remained textually stable across iterations, and items that were
removed, reworded, or introduced later were interpreted as evidence
situated within the period in which they were present.

The fixed-denominator convention, although it produces conservative
estimates, is fully replicable: any researcher with access to the
response spreadsheet can reproduce the reported percentages using the
same calculation rule.

The qualitative analysis combined deductive and inductive strategies
following Braun and Clarke~\cite{braun2006thematic}.
The interpretation of codes, categories, and themes may be influenced
by the researchers.
To increase traceability, records were kept of the transcripts,
assigned codes, provisional categories, script versions, and
instrument refinement decisions made throughout the study.

\section{Conclusion}
\label{sec:conclusion}

This study investigated the factors that affect software developers'
perceived efficiency over time, within a professional training context
in which the organization responsible for the monitoring had no
employment relationship with, nor operational control over, the
participants.
Monitoring 27 developers across twelve data collection waves, combining
recurring surveys and semi-structured interviews, made it possible to
observe factor behavior patterns that cross-sectional assessments
cannot reveal.

In response to RQ1, the most frequent factors were organizational in
nature: dependency on other areas or teams and waiting for external
responses or validations were present in nearly all waves, with means
above 37\% and 44\%, respectively.
Technical barriers, such as lack of technical knowledge and time spent
searching for information, were also frequent, especially in the early
waves.

In response to RQ2, the factors did not show uniform behavior over
time.
Organizational dependencies remained structural barriers throughout the
monitoring period, regardless of developers' accumulated experience.
Technical barriers showed a progressive reduction in the intermediate
waves, compatible with a process of adaptation to the technical
context, but returned in the final weeks, indicating that the
adaptation is not permanent.
The analysis also showed that part of the apparent reduction observed
in the more persistent factors is an artifact of lower response rates
in specific weeks, rather than a genuine resolution of the barriers.

In response to RQ3, the thematic analysis of the 18 interviews
identified seven themes that converged with the survey factors:
dependencies and communication, understanding of requirements and the
system, processes and planning, technical knowledge, interruptions and
focus, quality and rework, and the use of AI tools as a source of
support and challenges.
The convergence between the two channels strengthens the evidence that
the identified barriers reflect real conditions of participants' work.
The AI use theme was the most recurrent one in the interviews and
emerged spontaneously in the qualitative data before any corresponding
closed item existed in the surveys, demonstrating that longitudinal
monitoring with an adaptive instrument can identify emergent dimensions
that would be missed by fixed instruments.

This study offers three contributions to Empirical Software
Engineering.
First, it presents longitudinal evidence that efficiency barriers
behave differently over time: organizational dependencies are
structural and persist; technical barriers are adaptive and fluctuate;
and new dimensions, such as AI use, can emerge during the monitoring
period itself.
Second, it documents that a cross-sectional assessment conducted at
different points in the study would produce opposite conclusions about
the severity of the same barriers, providing direct empirical evidence
of the risks of one-off assessments in software development contexts.
Third, it demonstrates that intermediate results from longitudinal
monitoring can inform concrete organizational decisions during the
execution of the study itself, as occurred with the creation of a
course focused on requirements analysis and improvement.

The study has limitations arising from the use of self-reported
perceptions, from being conducted in a single context, from
intermittent non-response across waves, and from the absence of a
control group.
The findings do not support causal inferences or statistical
generalization to other groups of developers.
The results regarding AI use are preliminary and were obtained across
only three data collection waves.

As future work, we recommend replicating the study in other
organizational contexts and software development domains, comparing
the behavior of the factors in organizations that directly employ
developers with consulting and professional education contexts, and
extending the observation period for factors related to AI tool use to
verify whether the observed patterns consolidate, intensify, or change
as developers accumulate experience with these tools.
Future studies could also incorporate perceived severity indicators to
distinguish rare and severe factors from frequent, lower-impact
factors, and use more robust strategies for handling missing data in
weeks with lower response rates.

Monitoring developer efficiency over time reveals that what looks like
an improvement may be temporary adaptation, and what looks stable may
be a structural problem that no individual intervention can resolve.
Distinguishing between these two patterns is what makes longitudinal
monitoring indispensable for organizations that need to make training,
mentoring, and intervention decisions based on evidence, rather than on
one-off assessments that capture only a single moment of a dynamic
trajectory.

\bibliographystyle{IEEEtran}
\bibliography{references}

@article{noda2023devex,
  author    = {Noda, Abi and Storey, Margaret-Anne and Forsgren, Nicole
               and Greiler, Michaela},
  title     = {{DevEx}: What Actually Drives Productivity},
  journal   = {{ACM} Queue},
  volume    = {21},
  number    = {2},
  pages     = {35--53},
  year      = {2023},
  doi       = {10.1145/3595878}
}

@book{forsgren2018accelerate,
  author    = {Forsgren, Nicole and Humble, Jez and Kim, Gene},
  title     = {Accelerate: The Science of Lean Software and DevOps},
  publisher = {IT Revolution Press},
  year      = {2018},
  address   = {Portland, OR}
}

@article{meyer2019,
  author  = {Meyer, Andr{\'e} N. and Barr, Earl T. and Bird, Christian and Zimmermann, Thomas},
  title   = {Today was a Good Day: The Daily Life of Software Developers},
  journal = {IEEE Transactions on Software Engineering},
  volume  = {47},
  number  = {5},
  pages   = {863--880},
  year    = {2019},
  doi     = {10.1109/TSE.2019.2904957}
}

@unpublished{ademm_metodo,
  author = {Ribeiro, Danilo and Alves, Breno and Souza, Gabriel and Fran{\c{c}}a, C{\'e}sar and Souza, Alberto},
  title  = {{ADEMM: A Longitudinal Method for Monitoring Developer Efficiency in Industry}},
  note   = {Em preparação},
  year   = {2026}
}

@article{space2021,
  author  = {Forsgren, Nicole and Storey, Margaret-Anne and Maddila, Chandra and
             Zimmermann, Thomas and Houck, Brian and Butler, Jenna},
  title   = {The {SPACE} of Developer Productivity: There Is More to It Than
             You Think},
  journal = {Queue},
  year    = {2021},
  volume  = {19},
  number  = {1},
  pages   = {20--48}
}

@inproceedings{meyer2014software,
  title={Software developers' perceptions of productivity},
  author={Meyer, Andr{\'e} N and Fritz, Thomas and Murphy, Gail C and Zimmermann, Thomas},
  booktitle={Proceedings of the 22nd ACM SIGSOFT international symposium on foundations of software engineering},
  pages={19--29},
  year={2014}
}

@inproceedings{coelho2025software,
  title={Software Developers’ Perceptions of Productivity: An Industry-focused Study},
  author={Coelho, Murilo and Reinbold, Isabelle and Sancho, Lizie and Paixao, Matheus and Ara{\'u}jo, Allysson Allex and Freire, S{\'a}vio},
  booktitle={Simp{\'o}sio Brasileiro de Qualidade de Software (SBQS)},
  pages={12--22},
  year={2025},
  organization={SBC}
}

@inproceedings{cheng2022,
  title={What improves developer productivity at google? code quality},
  author={Cheng, Lan and Murphy-Hill, Emerson and Canning, Mark and Jaspan, Ciera and Green, Collin and Knight, Andrea and Zhang, Nan and Kammer, Elizabeth},
  booktitle={Proceedings of the 30th ACM Joint European Software Engineering Conference and Symposium on the Foundations of Software Engineering},
  pages={1302--1313},
  year={2022}
}

@article{razzaq2024systematic,
  title={A systematic literature review on the influence of enhanced developer experience on developers' productivity: Factors, practices, and recommendations},
  author={Razzaq, Abdul and Buckley, Jim and Lai, Qin and Yu, Tingting and Botterweck, Goetz},
  journal={ACM Computing Surveys},
  volume={57},
  number={1},
  pages={1--46},
  year={2024},
  publisher={ACM New York, NY}
}

@article{d2024measuring,
  title={Measuring developer experience with a longitudinal survey},
  author={D’Angelo, Sarah and Lin, Jessica and Dicker, Jill and Egelman, Carolyn and Hodges, Maggie and Green, Collin and Jaspan, Ciera},
  journal={IEEE software},
  volume={41},
  number={4},
  pages={19--24},
  year={2024},
  publisher={IEEE}
}

@inproceedings{cheng2022improves,
  author    = {Cheng, Lan and Murphy-Hill, Emerson and Canning, Mark
               and Jaspan, Ciera and Green, Collin and Knight, Andrea
               and Zhang, Nan and Kammer, Elizabeth},
  title     = {What Improves Developer Productivity at {Google}?
               {Code} Quality},
  booktitle = {Proceedings of the 30th ACM Joint European Software
               Engineering Conference and Symposium on the Foundations
               of Software Engineering (ESEC/FSE)},
  pages     = {1302--1313},
  year      = {2022},
  doi       = {10.1145/3540250.3558940}
}

@article{chapetta2020towards,
  title={Towards an evidence-based theoretical framework on factors influencing the software development productivity},
  author={Chapetta, Wladmir Araujo and Travassos, Guilherme Horta},
  journal={Empirical Software Engineering},
  volume={25},
  number={5},
  pages={3501--3543},
  year={2020},
  publisher={Springer}
}

@inproceedings{canedo2019factors,
  title={Factors affecting software development productivity: An empirical study},
  author={Canedo, Edna Dias and Santos, Giovanni Almeida},
  booktitle={Proceedings of the XXXIII Brazilian Symposium on Software Engineering},
  pages={307--316},
  year={2019}
}

@inproceedings{coutinho2024role,
  title={The role of generative ai in software development productivity: A pilot case study},
  author={Coutinho, Mariana and Marques, Lorena and Santos, Anderson and Dahia, Marcio and Fran{\c{c}}a, Cesar and de Souza Santos, Ronnie},
  booktitle={Proceedings of the 1st ACM International Conference on AI-Powered Software},
  pages={131--138},
  year={2024}
}

@book{wohlin2012experimentation,
  title={Experimentation in software engineering},
  author={Wohlin, Claes and Runeson, Per and H{\"o}st, Martin and Ohlsson, Magnus C and Regnell, Bj{\"o}rn and Wessl{\'e}n, Anders and others},
  volume={236},
  year={2012},
  publisher={Springer}
}

@article{braun2006thematic,
  author  = {Braun, Virginia and Clarke, Victoria},
  title   = {Using Thematic Analysis in Psychology},
  journal = {Qualitative Research in Psychology},
  volume  = {3},
  number  = {2},
  pages   = {77--101},
  year    = {2006},
  doi     = {10.1191/1478088706qp063oa}
}

@article{lynn2009methods,
  title={Methods for longitudinal surveys},
  author={Lynn, Peter},
  journal={Methodology of longitudinal surveys},
  pages={1--19},
  year={2009},
  publisher={Wiley Online Library}
}

@article{becker2025measuring,
  title={Measuring the impact of early-2025 AI on experienced open-source developer productivity},
  author={Becker, Joel and Rush, Nate and Barnes, Elizabeth and Rein, David},
  journal={arXiv preprint arXiv:2507.09089},
  year={2025}
}

@article{russo2024developers,
  author  = {Russo, Daniel and Hanel, Paul H. P. and van Berkel, Niels},
  title   = {Understanding Developers Well-Being and Productivity: A 2-year Longitudinal Analysis during the COVID-19 Pandemic},
  journal = {ACM Transactions on Software Engineering and Methodology},
  volume  = {33},
  number  = {3},
  pages   = {57:1--57:44},
  year    = {2024},
  doi     = {10.1145/3638244}
}

@article{kuutila2021individual,
  author  = {Kuutila, Miikka and M{\"a}ntyl{\"a}, Mika and Claes, Ma{\"e}lick and Elovainio, Marko and Adams, Bram},
  title   = {Individual Differences Limit Predicting Well-Being and Productivity Using Software Repositories: A Longitudinal Industrial Study},
  journal = {Empirical Software Engineering},
  volume  = {26},
  number  = {5},
  pages   = {88},
  year    = {2021},
  doi     = {10.1007/s10664-021-09977-1}
}

\end{document}